\documentstyle[12pt,aaspp4]{article}

\begin{document}

\oddsidemargin  -0.5pc
\evensidemargin -0.5pc

\def\dots{$\ldots$}
\def\p{\partial}
\def\bigskip{\vskip1true cm}
\def\prop{\propto}
\def\kms{km~s$^{-1}$}
\def\cm3{cm$^{-3}$}

\title{An Axisymmetric, Radiative Bow Shock Model with a Realistic Treatment of Ionization and Cooling}

\author{Alejandro C. Raga\altaffilmark{1}, Garrelt Mellema\altaffilmark{2} and Peter Lundqvist\altaffilmark{3}}

\altaffiltext{1}{Instituto de Astronom\'\i a, UNAM, Ap.~Postal~70-264, 04510 M\'exico, D. F., M\'exico. E-mail: raga@astroscu.unam.mx}
\altaffiltext{2}{Stockholm Observatory, S-133 36 Saltsj\"obaden, Sweden. E-mail: garrelt@astro.su.se}
\altaffiltext{3}{Stockholm Observatory, S-133 36 Saltsj\"obaden, Sweden. E-mail: peter@astro.su.se}

\begin{abstract}
We have chosen a reduced set of 18 ionization rate equations (for ions
of H, C, N, O, S and Ne), which allow us to obtain a moderately
accurate estimate of the non-equilibrium radiative cooling function.
We evaluate the accuracy of this approach by comparing our cooling
function with previous calculations, computed with a more extended set
of ions, for the case of gas that cools from a high temperature at
constant density. We also compute steady, plane shock models, which we
find to compare well with models calculated with much more detailed
microphysics.

Using our reduced set of rate equations, we present a simulation of a
radiative bow shock formed by a supersonic, plane stream impinging
on a rigid sphere.  The parameters for the calculation are chosen as to
approximately represent typical values found for Herbig-Haro objects,
and to give a cooling distance to bow shock radius ratio of 1/10.
This simulation is done with an adaptive grid code, which allows a
reasonably high resolution (with $\geq 25$ points) of the standoff
distance between the bow shock and the rigid obstacle.

Contrary to the standard expectation, we find that the bow shock standoff
distance in the stagnation region is considerably shorter than the
cooling distance behind a plane-parallel shock. Also, the centrifugal
pressure is found to play an important role in determining the
structure of the recombination region. This appears to
partially invalidate previous bow shock models based on a ``quasi-1D''
approach, at least for the particular parameters chosen for the
present simulation.

Finally, we present tabulations of the cooling rates that have been
used (for the different species), in order to facilitate the inclusion
of this treatment in other gasdynamic codes.
\end{abstract}


\section{Introduction}

Both axisymmetric and three-dimensional (3D) numerical simulations of
Herbig-Haro (HH) flows have been carried out during the past 10
years. These flows are more difficult to compute than extragalactic
jets because of the importance of the radiative cooling for the
dynamics of HH jets. The cooling distances are usually short, which
makes them difficult to resolve in numerical simulations using
standard methods.  It is even more difficult to predict the
observational characteristics of the computed flows, since the line
emission depends on the detailed density, temperature and
non-equilibrium ionization structure of the post-shock gas.

The axisymmetric and 3D simulations found in the literature are based
on different approximations of the ionization state and cooling rate
of the flow:

\begin{itemize}

\item[(i)] Raga \& B\"ohm (1987), Raga et al. (1988) and Raga (1988)
assumed that the ionization state of the gas is given by the coronal
ionization equilibrium condition (which is clearly incorrect for the
post-shock cooling regions), and used a cooling rate consistent with
this assumption.

\item[(ii)] Blondin, K\"onigl, \& Fryxell (1989) and Blondin, Fryxell, \&
K\"onigl (1990) used a ``non-equilibrium'' cooling function, which was
computed by Kafatos (1973) for a parcel cooling at constant density
from a high temperature. While this cooling rate might mimic some of
the non-equilibrium ionization effects found in a recombination region
of a shock wave, it is unclear to what extent it departs from a more
consistently computed cooling function. Also, these authors quite
surprisingly assumed that the gas consists of fully ionized hydrogen
(regardless of position), not including the effect of partial
ionization on the gas pressure. This assumption leads to a very high
cooling rate at low temperatures. To avoid this Blondin et al. 
introduced an artificial cutoff temperature of $\sim 10^4$~K below
which the cooling function was set to zero. An apparently almost
identical approach was taken by Gouveia dal Pino \& Benz (1993, 1994)
and Chernin et al. (1994).

\item[(iii)] Stone \& Norman (1993ab, 1994ab)
explicitly computed the non-equilibrium ionization state of hydrogen
(integrating a single ionization rate equation), and used this to
obtain the electron density and the densities of the ``ionized'' and
``neutral'' components of the gas. These densities were then used to
compute an approximate cooling rate. It appears that Stone \& Norman
neglected cooling due to collisional ionization of hydrogen, leading
to an unrealistically high state of ionization behind shocks with
velocities $\leq 100$~\kms (moving into mostly neutral gas). An
equivalent approach was taken by Raga (1994), Falle \& Raga (1995),
Biro, Raga, \& Cant\'o (1995), though these authors included cooling
due to collisional ionization of hydrogen, and in some cases (Raga et al.
1995) also cooling due to the presence of H$_2$ molecules.

\end{itemize}

The axisymmetric, nonadiabatic simulations made previously by
R\`o\.zyczka \& Tenorio-Tagle (1985) apparently used an approach
similar to the ones described in (i) or (ii) However, they
do not give an extensive description of their treatment of the
cooling.

While these three approaches result in a more or less correct description
of the dynamical characteristics of the flow, their approximate treatment
of the ionization and cooling is likely to lead to unrealistic
predictions of the emitted spectrum.  Also, the numerical resolution
of the post-shock flow in these calculations is too coarse to obtain
reliable predictions of the emitted spectrum.
The lack of resolution can to some
extent be solved by computing only a limited region of an HH flow (as
in the bow shock models of Raga \& B\"ohm 1987) and/or using an
adaptive computational grid as done by Falle \& Raga (1993, 1995) and
Raga (1994).

Because of these limitations, most of the detailed comparisons between
model predictions and observations of HH objects have been based on
applications of ``quasi-1D'' bow shock models (e.g., Hartmann \&
Raymond 1984; Choe, B\"ohm, \& Solf 1985; Raga \& B\"ohm 1985, 1986;
Hartigan, Raymond, \& Hartmann 1987; Hartigan, Raymond, \& Meaburn
1990). In these models, a surface that approximately reproduces the
shape of the bow shock is considered, and the emission from the
post-shock region is computed by assuming that the surface locally has
an emission identical to the one from a plane, steady shock of the
corresponding shock velocity.  These plane shocks allow a detailed
treatment of ionization, cooling and radiative transfer effects (see,
e.g., Raymond 1979; Hartigan et al. 1987).  It is thought (though not
proven) that this approximate approach is valid for flow parameters
that produce cooling distances which are much shorter than the radius
of the bow shock. As it is unclear to what extent these quasi-1D
models are realistic approximations of a bow shock, it is of obvious
interest to calculate an axisymmetric, physically more realistic
simulation of the flow. This is now computationally possible.

An example of this is the method for including non-equilibrium cooling
in a gas dynamics code described by Frank \&\ Mellema (1994a). This
code was used to study the formation and evolution of (largely)
photo-ionized nebulae, such as Planetary Nebulae (Frank \&\ Mellema
1994b, Mellema \&\ Frank 1995, Mellema 1995). The success of this
approach was one of the motivations for the current work. An important
difference between their and our method (apart from the
photo-ionization) is that they used analytical fits to the different
cooling rates and atomic rates (see Balick, Mellema, \&\ Frank 1993),
whereas we use look-up tables which include more processes.

Another modern development helpful in studying the gas dynamics of
problems with short cooling distances is the adaptive grid
approach. An adaptive grid provides a huge improvement in both memory
and computing time requirements, and makes it possible to carry out
numerical simulations which include an extensive set of microphysical
processes (e.g., rate equations describing ionization and/or chemical
processes for a number of species). Clearly, such an approach results
in a much more realistic description of both the flow and its
observational characteristics.

We have selected a set of 18 ionization rate equations for the species
necessary to obtain realistic cooling rates for shock velocities up to
$\sim 200$~\kms.  This paper presents a detailed description of our
ionization and cooling rate calculations, as well as a quantitative
evaluation of the resulting cooling function (\S 2). In \S 3
we then present results for a high-resolution, adaptive grid
simulation of a bow shock formed around a rigid sphere, in which we
have used our description of the ionization and cooling.  The results
from this simulation are used to illustrate the detailed
characteristics of a radiative bow shock flow, and to compare with the
results from a ``quasi-1D'' model.  Finally, in Appendix~A we present
details of the ionization, recombination and cooling rates which we
have used, including tables for the cooling rates.  This should
facilitate implementations of this network into other gas dynamic
codes.

\section{A simplified ionization and cooling rate network}

We consider a reduced set of rate equations describing the
time-evolution of the non-equilibrium ionization fractions of H~II,
C~III-IV, N~II-V, O~II-V, S~III-V and Ne~II-V. We assume that C and S
are always at least singly ionized, and complete this set of rate
equations with the conservation condition for all of the ions of each
element. The processes we include are collisional ionization,
radiative and dielectronic recombination, and charge exchange
processes of different ions with H and H$^+$. A more detailed
description of this is given in Appendix~A.

The rate equations for the ions can be integrated in step with the
appropriate dynamical equations. The resulting ion and electron
densities can then be used to calculate the cooling rate $L$ (energy
per unit volume and time). In our prescription of $L$, we include
hydrogen collisional excitation of Lyman-$\alpha$, as well as
radiative recombination and free-free emission. For the other ions we
include collisional excitation of optical and ultraviolet lines.  A
complete tabulation of the cooling rate $L_{iz}/(n_e n_{iz})$ for each
ion of species $i$ and charge $z$ is given in Tables~1-11 (see
Appendix~A).

To test the accuracy of our reduced set of ions and the resulting
cooling function, we computed the evolution of a parcel that cools at
constant density from a high initial temperature. We have considered a
parcel with a number density (atoms plus ions) $n=1$~\cm3, cooling
from an initial temperature $T_1=10^6$~K, and recombining from coronal
ionization
equilibrium evaluated at this temperature. The same problem was
studied by Innes (1985), who, however, considered a much larger set of
ions and lines.

Figure~1 shows a comparison between the time-dependent cooling
function using our reduced set of ionization rate equations, and the
more complete calculation of Innes (1985). At temperatures above
$2\times 10^5$~K our cooling function is lower than the one of Innes
(1985) by a factor of $\sim 2$. This is because our set of equations
extends only to ions of ionization stage V. One more ionization stage
is necessary to compute a model that starts at $T_1=10^6$~K. However,
below $2\times 10^5$~K, our simplified prescription results in a
cooling function that deviates by less than 20~\% from Innes (1985).

Another interesting check of our reduced set of equations is to
compute models of steady, planar shocks.  We included the cooling
function in the differential equation for the evolution of the
enthalpy, which is integrated simultaneously with the ionization rate
equations (see, e.g., Raymond 1979). We have computed a set of models
with a pre-shock
density $n_0=100$~\cm3, and shock velocities $v_s=40,$ 60, 80, 100,
120, 140, 160 and 180~\kms. We have computed this set of models twice,
one time for the case of a neutral pre-shock gas (except for C and S, which
are always at least singly ionized, see above), and another for singly
ionized H and O.

To compare with previous, more detailed shock wave models, we computed
the distances at which the post-shock gas has cooled to $10^4$~K and
$10^3$~K, $d_4$ and $d_3$ respectively. We plot these cooling
distances as a function of shock velocity $v_s$ in Figure~2.

In this figure, we have also plotted $d_4$ and $d_3$ for the shock
models of Hartigan et al. (1987). These authors considered a very
extensive number of processes, and included a detailed treatment of
the radiative transfer of ionizing radiation. Figure 2 includes the
values of $d_4$ and $d_3$ for the two sets of shock models tabulated
by Hartigan et al. (1987). One of these corresponds to
``self-consistent preionization'' calculations in which the ionization
state of the pre-shock gas is set in a self-consistent way by the
ionizing radiation from the post-shock gas. In the second set (``fully
preionized'' models), it is assumed that the pre-shock gas has fully
ionized hydrogen.

From Figure~2, we see that the $d_4$ cooling distance calculated from
our ``fully preionized'' models agrees very well for all shock
velocities with the values obtained from the corresponding models of
Hartigan et al. (1987). A reasonably good agreement is also obtained
for the $d_3$ cooling distance, except for $v_s=40$~\kms. For this
velocity there is a discrepancy of a factor $\sim 3$.

We find (see Fig. 2) that our ``neutral pre-shock gas'' models
predict values of $d_4$ that are in reasonably good agreement with the
values obtained from the ``self-consistent pre-ionization'' models of
Hartigan et al. (1987) for shock velocities $v_s\geq 80$~\kms.
Deviations by a factor of $\sim 4$ are found for the two models with
lower velocity. A reasonably good agreement is also found between the
values of $d_3$ predicted from the shock models with $v_s\geq
100$~\kms, though discrepancies of a factor of $\sim 10$ are found for
lower velocities.

From this it is clear that our reduced set of ionization rate
equations produces shock models that are quite similar to the much
more detailed calculations of Hartigan et al. (1987).  The main
deviations occur in the low temperature tail of the recombination
region. Comparing the $d_4$ cooling distance with other codes
(which all have more detailed
treatments of ionization and cooling than ours), we find that our
models appear to fall well within the quite broad range of predictions
produced by these codes, as is clear from a comparison with the tabulations
of P\'equignot (1986) and Ferland et al. (1995). As these tabulations of
shock models do not include the $d_3$ cooling distance, it is not
clear whether the discrepancies in $d_3$ between our code and
the one of Hartigan et al. (1987) are particularly extreme (see above).

The main drawback of our approach is that we do not consider the radiative
transfer of ionizing radiation. This limits us to treat
pre-ionization as a free parameter, which is a shortcoming especially
at shock velocities around $v_s \approx 80$-100~\kms (see, e.g.,
Shull and McKee 1979). However, it is at the present stage premature
to include radiative transfer of the diffuse ionizing radiation field
in axisymmetric or 3D numerical simulations. This should be the
objective for future simulations. Here we mainly test our reduced set
of equations, which is a first step towards calculations which include
proper radiative transfer.

\section{A simulation of a radiative, blunt-body flow}

We have implemented our ionization and cooling rate prescription in
the axisymmetric, adaptive grid code Coral. Coral has been used quite
extensively for computing HH flows, albeit with a more simplified
treatment of the microphysical processes (see, e.g., Raga et al. 1995).

In the present configuration, a set of 23 equations is integrated: the
4 gasdynamic equations, 18 equations for the atomic and ionic species,
and 1 equation for an inert dye (which is used to trace material
surfaces). This represents an increase by a factor of $\sim 3$-5 in
number of equations with respect to all previous numerical simulations
of HH flows in the literature (see \S 1).  With our formalism, we
can calculate more realistic models of HH flows, especially in terms
of the relevant atomic processes (see \S 2).  As an illustration of
this capability, we have chosen to compute a relatively simple,
blunt-body flow, such as the one computed in the past by Raga \&
B\"ohm (1987).

The supersonic, blunt-body flow problem is not a very realistic
description of the flow expected in the ISM, but it has the clear
advantage that a single, main shock (i.e., the bow shock) is
produced. This simplicity allows a more detailed comparison with the
results from planar shock and quasi-1D bow shock models. In future
papers, we will present simulations of more complex flows (e.g., the
working surface of a jet).

We have considered a rigid, spherical obstacle of radius
$R_c=10^{16}$~cm interacting with a plane-parallel incident stream of
gas of velocity $v_0=80$~\kms~and temperature $T_0=1000$~K. This value
for the velocity was chosen to be well under the limit of
100-120~\kms~ above which the post-shock region would become thermally
unstable (see, e.g., Falle 1981; Raga \& B\"ohm 1987). The incident
stream is assumed to have a hydrogen ionization fraction of $10^{-3}$,
and a number density of $n_0=2.63$~\cm3.  For a steady, plane shock
model with these parameters, one obtains a cooling distance
$d_4=10^{15}$~cm$=R_c/10$.

The calculation is started with the plane flow in contact with the
surface of the obstacle.  The system of equations is first integrated
up to a time $t_1=65$~yrs in a domain of extent of $1.5\times
10^{16}$~cm (in both the axial and radial directions) with a
5-level, binary adaptive (cylindrical) grid (see Raga et al. 1995)
of minimum grid spacing $\Delta = 5.86\times 10^{13}$~cm (in both
axial and radial directions, which would correspond to a uniform grid
of $256\times 256$ points for the chosen grid size).  Then, the
maximum resolution of the grid is doubled, so that $\Delta=2.93\times
10^{13}$~cm, and the integration is continued until
$t_2=85$~yrs. Finally, the resolution is doubled again, resulting in a
minimum grid spacing $\Delta=1.46\times 10^{13}$~cm, and the numerical
integration is continued until $t_3=105$~yrs. The final flow (which we
discuss in detail below) is therefore computed in a grid that
corresponds to a uniform grid of $1024\times 1024$ points.  The
adaptive grid at the end of the integration, however, has a filling
factor of only 12~\% (with respect to a uniform grid).

The temperature, density and pressure distributions of this numerical
integration are shown in Figure 3, along with a representation of the
adaptive grid. One of the most striking features is that the flow is
non-steady (this is apparent from Fig.~3, even though this figure only
shows a single snapshot). This result may at first seem surprising,
since the flow parameters were chosen so as to produce a thermally
stable bow shock (see above). However, a more thorough check shows
that the non-steady features of the flow appear to be associated with
a cool (1.0-1.5$\times 10^4$~K), dense layer that is formed around the
head of the obstacle. The material in this dense layer flows outwards
along the surface of the obstacle, and detaches from the obstacle when
the cylindrical radius is $r_d\approx 0.8$-0.9~$R_c$.  The region
between the detached layer and the obstacle is filled with low
density, cool material. We speculate that the time-dependent features
of the bow shock might be associated with ``thin shell instabilities''
developing in this dense, cool layer (Dgani, Van Buren, \&
Noriega-Crespo 1996).

Another result that may seem surprising is that the standoff distance
of the stagnation region is only $\approx 4.7\times 10^{14}$~cm, which
is roughly equal to 1/2 of the cooling distance $d_4=10^{15}$~cm for a
steady, plane shock of the same pre-shock parameters. This goes
against the standard expectation that the standoff distance should be
approximately equal to $d_4$.

However, a close look at the structure of the recombination region
behind an 80~\kms, plane shock (as computed, e.g., by Raga 1989) shows
that the $d_4$ cooling distance does not correspond to a point where a
drastic compression occurs in the post-shock flow (as it does for
shock velocities $\geq 120$~\kms). Actually, the largest compression
occurs in the region where hydrogen is collisionally ionized right
after the shock. As $d_4$ therefore does not have any clear dynamical
significance, it is not surprising that the standoff distance does not
have the same value as the cooling distance.

This has quite interesting implications for the application of
``quasi-1D'' bow shock models (in which the post-bow shock emission is
approximated with locally one-dimensional recombination regions, see
\S 1).  From the above discussion we would conclude that it is
not possible to model bow shocks of velocities $\leq 100$~kms with
quasi-1D models, even when the cooling distance is much smaller than
the bow shock radius. This is because the standoff distance is
considerably smaller than the cooling distance.  A substantial part of
the emission is produced in regions where the streamlines are heavily
curved, producing both dynamical and thermal structures which are
different from the ones in a 1D shock model.

This can be quantified to some extent by analysing cuts across the
bow shock structure obtained from our numerical simulation. As shown
in Figure~4, we have chosen three cuts ($a$, $b$ and $c$) across the
flow, taken perpendicular to the bow shock at different distances from
the symmetry axis.

The flow along the on-axis cut ($a$, see Fig. 4) is shown in
Figures~5 and 6. Figure~5 shows the temperature, number density and
pressure, plotted as a function of a coordinate $l$, which is measured
from the bow shock along the direction of the cut.  The ordinate of
the plot extends up to the position of the rigid body.  Figure~6 shows
the ionization fractions of H~II, O~II and O~III along the same
cut. Also shown in Figures~5 and 6 are the corresponding variables
predicted from a steady, plane shock of the same pre-shock parameters.

If we compare the structure of the plane shock with the one measured
along cut $a$ (Figs. 4, 5 and 6), it is evident that the differences
are not very large. The small observed differences
could at least partially be due to the lower
resolution of the axisymmetric simulation, resulting in inaccuracies
in the computed structure of the cooling region.
From this, we would conclude that the on-axis structure of the bow
shock approximately corresponds to a plane shock model which is
``truncated'' at $l \sim d_4/2$.

The flow and ionization structure across cut $b$ (see Fig. 4) are
shown in Figures~7 and 8. These also include the corresponding
variables as predicted from a plane shock of velocity $v_s=66.4$~\kms
(which is the velocity that corresponds to the projection of the
preshock velocity normal to the local bow shock surface). The other
preshock parameters are identical to the ones of the bow shock model.
Differences between the axisymmetric simulation and the plane shock
model are now much more obvious than for cut $a$.  At $l\approx
10^{14}$~cm, the post-bow shock divergence of the streamlines results
in densities that are substantially lower than the ones of a plane
shock model (see Fig. 7). At $l\approx 10^{15}$~cm, we start to see
more highly ionized material (see Fig. 8). This is gas which has
gone through the stronger shocks found closer to the symmetry axis and
has then moved away towards the bow shock wings.

Finally, the flow and ionization structure across cut $c$ (see Fig.
4) are shown in Figures~9 and 10. Here we also show the structure of a
plane shock of velocity $v_s=40.2$~\kms (which corresponds to
the projection of the preshock velocity normal to the local bow shock
surface). We now find an almost complete lack of correspondence
between the structures predicted from the numerical simulation and
from the bow shock models, except immediately behind the shock. The
bow shock model has a dramatically higher ionization than the plane
shock at distances $l\sim 1.0$-$1.7\times 10^{15}$~cm (Fig.
10). This partially ionized region has temperatures ranging between
1000-12000~K (Fig. 9), and corresponds to gas that has gone through
stronger shocks closer to the symmetry axis. This gas has a ``frozen
in'' ionization state as a result of the adiabatic expansion that it
has undergone in going around the rigid obstacle.

An interesting feature seen in cut $b$, and even more strongly in cut
$c$, is the presence of a clear pressure gradient in the
post-bow shock region (see Figs. 7 and 9), which is absent in the
structures found from all plane shock models. This pressure gradient
is due to the so-called ``centrifugal effect'', and is necessary for
forcing the postshock material to follow a curved trajectory, along
the surface of the obstacle.

It should be pointed out again that our simulation has the important
limitation of corresponding to a bow shock formed around a rigid body
(which is of course not very realistic for the case of a flow in the
ISM). However, this simulation would be directly applicable for describing
the early stages of the interaction of a high velocity, low
density wind with a dense, spherical cloud. The evolution of this
interaction, however, will lead to a clear deformation of the cloud
(and to complex effects involving mixing between the wind and cloud
material). All of these effects will clearly lead to more major departures
(than the ones described above)
from the simple picture invoked in a quasi-1D bow shock model.

\section{Discussion}

In this paper, we have presented a description of a reduced set of
ionization rate equations, which has been devised for incorporating
into 2D or 3D gasdynamic simulations of HH flows, but parts of which
might be applied to other astrophysical flows as well (PNe, SN,
etc.). Through a comparison with the time-dependent cooling function
of Innes (1985), we find that the cooling function computed using only
the contributions from our reduced set of ions is acceptably accurate
in the $10^4-10^6$~K temperature range.

We then compute steady, plane shock models with this ionization rate
equation set, and compare the cooling distances found for shocks of
velocities $v_s=40$ to 180~\kms\ with the corresponding predictions
from the much more detailed models of Hartigan et al. (1987). We again
find a good agreement. From this comparison, we also find that the
main limitation of our approach is that we do not include radiative
transfer of the ionizing radiation. Unfortunately, this is crucial for
determining the amount of preionization of the gas that enters the shock
wave. At present we are forced to treat the preionization as a free
parameter. As it is quite complex to incorporate radiative transfer in
2D or 3D gasdynamic simulations, we have not yet attempted to carry
out this important step. This is left for a future study.

Finally, we present an axisymmetric simulation of a bow shock formed
around a rigid obstacle. This problem was chosen because of the
relative simplicity of the resulting shock structure, and also because
of its relevance for HH flows. Even though the computed model has a
relatively short cooling distance of 1/10 of the radius of the
obstacle, we find that the recombination region behind the bow shock
deviates quite radically from what is predicted from ``quasi-1D''
models, based on a superposition of oblique, 1D shock models. 

Although this result cannot be directly applied to HH objects (which
are not rigid obstacles), it does show that quasi-1D bow shock
models are not really appropriate for describing flows of this type.
For example, we
find shorter standoff distances and higher ionization conditions in the
wings than in these types of models. This would suggest that the
results of the analysis of HH spectra using quasi-1D bow shock models
should be used with some caution.

In order to carry out comparisons of our models with real HH objects we
need to compute jet models of the type described in \S 1. With
the details from the non-equilibrium ionization calculations, it will
be possible to present the ionization stratification distribution, and
using the velocity information, line profiles. In a second paper we will
present the results of such an axisymmetric simulation and a detailed
comparison with observations of HH objects.

\acknowledgments

GM thanks the IAUNAM for a very pleasant stay in Mexico City, during
which the bulk of this work was carried out. GM's stay at the
IAUNAM as well as part of the work of ACR was supported
by the DGAPA (UNAM) grant IN~105295 and the
UNAM-Cray grant SC-004596. ACR and GM thank Jorge
Cant\'o, Will Henney, and Jane Arthur for useful discussions. The 
research of PL is supported by the Swedish Natural Science Research
Council.

\appendix
\section{Appendix A}

As discussed in \S 2, we have included the ions H~I-II, C~II-IV,
N~I-V, O~I-V, Ne~I-V and S~II-V in the ionization rate equations and
the cooling function. These ions constitute the major part of the
cooling for temperatures of up to a few $\times 10^5$~K, except for
sulphur which was included because of its importance for
nebular analysis.  The abundances (by number) we have used are
$x_H=0.999$, $x_C=3.3\times 10^{-4}$, $x_O=6.6\times 10^{-4}$,
$x_{Ne}=8.3\times 10^{-4}$ and $x_S=1.6\times 10^{-5}$.  Except for
hydrogen, these abundances correspond to average cosmic values.
Helium was excluded since it is uninteresting for diagnostic purposes,
and does not have a strong effect on the cooling. Obviously, this
omittance also speeds up the calculations slightly. In our equation of
state, we have assumed that the hydrogen atoms have an increased mass
of $1.3\times m_H$. This is equivalent to including helium in the
equation of state assuming an He/H-ratio (by number) of 0.10. However,
we do not take into account the fact that helium and hydrogen have
different ionization potentials.

In the ionization rate equations we have considered collisional
ionization, radiative and dielectronic recombination, as well as
charge exchange reactions with hydrogen. Data for these processes, as
well as the cooling discussed below, overlap to a large extent with
the more complete data set used by Lundqvist \& Fransson (1996) and
Lundqvist et al. (1996). However, for the sake of completeness we
discuss the atomic data here too.

For hydrogen we have used the Case A recombination coefficient of
Seaton (1959), and the collisional ionization coefficient
from Cox (1970). For some of the ions (C II, N II,
O III, S II and S III) we have used the complete recombination rates
(radiative + dielectronic) of Nahar (1995). For the other ions we have
used a combination of different radiative and dielectronic rates. For
recombination to N I, N III, O II, O IV, Ne I-IV and S IV we have
included the radiative rates of Landini \& Monsignori Fossi (1990),
for recombination to C III and N IV those of Arnaud \& Rothenflug
(1985), and for recombination to O I that of Chung, Lin, \& Lee
(1991). Low-temperature dielectronic rates for recombination to C III,
N I, N IV and O II were taken from Nussbaumer \& Storey (1983), to Ne
II-IV from Nussbaumer \& Storey (1987), whereas for S IV we have used
the approximate rate from Lundqvist \& Fransson (1991). In addition to
this, we have included the high-temperature rates of Landini \&
Monsignori Fossi (1990) for recombination to C III, N I, N IV, O II
and Ne I-IV, the rates coveraging a wider temperature range calculated
by Badnell (1987, 1991) for recombination to N III and S II-IV, and by
Badnell \& Pindzola (1989) for recombination to O I-II and O
IV. Collisional ionization of C, N and O was taken from Landini \&
Monsignori Fossi (1990) with slight adjustment to agree better with
the results of Lotz (1967).  For Ne and S the detailed fits of Arnaud
\& Rothenflug (1985) were used. For charge transfer reactions with
hydrogen we have considered the same rates as in the comprehensive
compilation of Kingdon \& Ferland (1996).

Hydrogen cooling was assumed to be due to collisional ionization and
collisional excitation of Ly-$\alpha$, using the collision strength of
Aggarwal (1983).  Metal cooling was calculated using multi-level models
for C II-III, N I-III, O I-IV, Ne III-V and S II-IV. For the other
ions we only considered the most important allowed and semi-forbidden
transitions. The resulting cooling rate per electron, and per ion of
species $i$ and charge $z$, $\Lambda_{i,z}=L_{i,z}/(n_{i,z}n_e)$, is
given in Tables 1-10 for C~I-III, O~I-IV and Ne~I-V for a wide range
of temperatures ($10^2$ -- $10^6$~K) and electron density (1.0 --
$10^6$~cm$^{-3}$).  (Note that we have included the C~I cooling in
these tables, even though it was not incorporated in our hydro
simulations).  The first column of the tables gives the values of
${\rm log}_{10} T_e$ (with $T_e$ in K), and the 13 following columns
give the values of ${\rm log}_{10} \Lambda_{i,z}$ (with
$\Lambda_{i,z}$ in erg~cm$^3$s$^{-1}$) for equidistant values of ${\rm
log}_{10} n_e$ (with $n_e$ in \cm3), respectively.

The atomic data we have used are to a fair extent from Gaetz \&
Salpeter (1983), Mendoza (1983), Gallagher \& Pradhan (1985), Aggarwal
et al. (1986) and Osterbrock (1989). We have added and updated data for
C II (Bi\'emont, Delahaye, \& Zeippen 1994; Peng \& Pradhan 1995), C
III (Allard et al. 1990; Keenan, Feibelman, \& Berrington 1992; Fleming,
Hibbert, \& Stafford 1994), N II (Stafford et al. 1994) N III
(Stafford, Bell, \& Hibbert 1994; Brage, Froese-Fischer, \& Judge 1995;
Peng \& Pradhan 1995), O II (McLaughlin \& Bell 1993), O III (Aggarwal
1993) and O IV (Zhang, Graziani, \& Pradhan 1994; Peng \& Pradhan
1995).  Although not included in our tables, the S II model ion
included the results of Cai \& Pradhan (1993) and Keenan et al. (1993).

As can be seen from Tables 1-10, $\Lambda_{i,z}$ decreases with
increasing density. This is due to collisional de-excitation which
suppresses forbidden transitions in the range of densities we
consider.  For some of the ions (H~I, C~IV and O~V), only allowed
transitions contribute.  These transitions are not affected by
collisional de-excitation in our density range (Table~11). We have
assumed that this also applies to the recombination emission of H~II
(cf. Table~11), which is a good approximation.

In calculating the cooling, we have not considered optical depth
effects in the lines. The main effect is that resonance
lines shortward of the Lyman edge of hydrogen may contribute to the
ionization of hydrogen elsewhere in the flow.
Considering the fact that we do not include helium
in our models, nor ionization due to the diffuse continuum emission,
this effect is unimportant.  At low temperatures the omission of
optical depth effects in low-lying fine structure lines can result in
an overestimate of the cooling, but probably not to the level of other
uncertainties in our calculations.


\clearpage

\begin{deluxetable}{llllllllllllll}
\scriptsize
\tablewidth{0pc}
\tablenum{3}
\tablecaption{C III cooling}
\tablehead{}
\startdata
 3.6 & -26.012 & -26.012 & -26.012 & -26.012 & -26.013 & -26.013 & -26.013 & -26.015 & -26.021 & -26.033 & -26.050 & -26.064 & -26.071 \\
 3.7 & -24.377 & -24.377 & -24.377 & -24.377 & -24.377 & -24.377 & -24.378 & -24.379 & -24.384 & -24.395 & -24.411 & -24.425 & -24.432 \\
 3.8 & -23.089 & -23.089 & -23.089 & -23.089 & -23.089 & -23.089 & -23.090 & -23.091 & -23.096 & -23.106 & -23.121 & -23.134 & -23.140 \\
 3.9 & -22.078 & -22.078 & -22.078 & -22.078 & -22.078 & -22.078 & -22.078 & -22.080 & -22.084 & -22.093 & -22.107 & -22.119 & -22.126 \\
 4.0 & -21.284 & -21.284 & -21.284 & -21.284 & -21.284 & -21.285 & -21.285 & -21.286 & -21.289 & -21.298 & -21.310 & -21.322 & -21.329 \\
 4.1 & -20.659 & -20.659 & -20.659 & -20.659 & -20.659 & -20.659 & -20.659 & -20.660 & -20.663 & -20.670 & -20.682 & -20.693 & -20.700 \\
 4.2 & -20.155 & -20.155 & -20.155 & -20.155 & -20.155 & -20.155 & -20.155 & -20.156 & -20.159 & -20.165 & -20.176 & -20.186 & -20.192 \\
 4.3 & -19.733 & -19.733 & -19.733 & -19.733 & -19.733 & -19.733 & -19.733 & -19.734 & -19.736 & -19.741 & -19.751 & -19.760 & -19.766 \\
\enddata
\end{deluxetable}
\clearpage

\begin{deluxetable}{llllllllllllll}
\scriptsize
\tablewidth{0pc}
\tablenum{4}
\tablecaption{O I cooling}
\tablehead{}
\startdata
 2.0 & -23.932 & -23.932 & -23.933 & -23.933 & -23.933 & -23.935 & -23.939 & -23.952 & -23.986 & -24.062 & -24.210 & -24.474 & -24.857 \\
 2.1 & -23.683 & -23.683 & -23.683 & -23.683 & -23.684 & -23.685 & -23.691 & -23.706 & -23.745 & -23.831 & -23.993 & -24.271 & -24.663 \\
 2.2 & -23.478 & -23.478 & -23.478 & -23.478 & -23.479 & -23.481 & -23.487 & -23.504 & -23.550 & -23.646 & -23.821 & -24.113 & -24.515 \\
 2.3 & -23.309 & -23.309 & -23.309 & -23.309 & -23.310 & -23.312 & -23.319 & -23.339 & -23.390 & -23.496 & -23.686 & -23.992 & -24.402 \\
 2.4 & -23.169 & -23.169 & -23.169 & -23.169 & -23.170 & -23.172 & -23.180 & -23.203 & -23.260 & -23.376 & -23.580 & -23.899 & -24.317 \\
 2.5 & -23.052 & -23.052 & -23.052 & -23.053 & -23.053 & -23.056 & -23.065 & -23.090 & -23.153 & -23.280 & -23.496 & -23.827 & -24.253 \\
 2.6 & -22.954 & -22.954 & -22.955 & -22.955 & -22.956 & -22.959 & -22.969 & -22.997 & -23.065 & -23.201 & -23.429 & -23.772 & -24.204 \\
 2.7 & -22.872 & -22.872 & -22.872 & -22.872 & -22.873 & -22.877 & -22.887 & -22.918 & -22.992 & -23.136 & -23.376 & -23.729 & -24.166 \\
\enddata
\end{deluxetable}
\clearpage

\begin{deluxetable}{llllllllllllll}
\scriptsize
\tablewidth{0pc}
\tablenum{5}
\tablecaption{O II cooling}
\tablehead{}
\startdata
 3.3 & -26.875 & -26.875 & -26.876 & -26.881 & -26.893 & -26.927 & -27.004 & -27.165 & -27.455 & -27.858 & -28.322 & -28.808 & -29.301 \\
 3.4 & -25.194 & -25.194 & -25.195 & -25.199 & -25.211 & -25.242 & -25.313 & -25.465 & -25.741 & -26.135 & -26.592 & -27.070 & -27.543 \\
 3.5 & -23.868 & -23.868 & -23.869 & -23.873 & -23.883 & -23.912 & -23.978 & -24.119 & -24.383 & -24.764 & -25.205 & -25.651 & -26.058 \\
 3.6 & -22.824 & -22.824 & -22.825 & -22.828 & -22.838 & -22.863 & -22.924 & -23.056 & -23.303 & -23.664 & -24.072 & -24.447 & -24.743 \\
 3.7 & -22.002 & -22.002 & -22.003 & -22.006 & -22.014 & -22.038 & -22.093 & -22.214 & -22.442 & -22.772 & -23.123 & -23.402 & -23.598 \\
 3.8 & -21.356 & -21.356 & -21.357 & -21.359 & -21.367 & -21.388 & -21.438 & -21.547 & -21.752 & -22.040 & -22.321 & -22.514 & -22.647 \\
 3.9 & -20.848 & -20.848 & -20.849 & -20.851 & -20.858 & -20.876 & -20.921 & -21.018 & -21.197 & -21.440 & -21.654 & -21.788 & -21.885 \\
 4.0 & -20.450 & -20.450 & -20.451 & -20.453 & -20.459 & -20.475 & -20.515 & -20.601 & -20.755 & -20.955 & -21.117 & -21.212 & -21.289 \\
\enddata
\end{deluxetable}
\clearpage

\begin{deluxetable}{llllllllllllll}
\scriptsize
\tablewidth{0pc}
\tablenum{6}
\tablecaption{O III cooling}
\tablehead{}
\startdata
 2.0 & -20.660 & -20.668 & -20.691 & -20.757 & -20.909 & -21.178 & -21.549 & -21.991 & -22.469 & -22.961 & -23.459 & -23.958 & -24.458 \\
 2.1 & -20.543 & -20.551 & -20.573 & -20.635 & -20.779 & -21.031 & -21.384 & -21.814 & -22.288 & -22.779 & -23.276 & -23.775 & -24.275 \\
 2.2 & -20.450 & -20.457 & -20.479 & -20.538 & -20.673 & -20.910 & -21.248 & -21.669 & -22.139 & -22.629 & -23.125 & -23.624 & -24.124 \\
 2.3 & -20.378 & -20.384 & -20.405 & -20.461 & -20.589 & -20.814 & -21.140 & -21.555 & -22.022 & -22.511 & -23.008 & -23.506 & -24.006 \\
 2.4 & -20.323 & -20.329 & -20.349 & -20.403 & -20.525 & -20.740 & -21.058 & -21.468 & -21.934 & -22.422 & -22.918 & -23.417 & -23.917 \\
 2.5 & -20.284 & -20.290 & -20.309 & -20.360 & -20.477 & -20.685 & -20.997 & -21.404 & -21.868 & -22.356 & -22.852 & -23.350 & -23.850 \\
 2.6 & -20.259 & -20.265 & -20.283 & -20.332 & -20.444 & -20.646 & -20.953 & -21.356 & -21.819 & -22.306 & -22.802 & -23.301 & -23.800 \\
 2.7 & -20.246 & -20.252 & -20.269 & -20.316 & -20.424 & -20.621 & -20.921 & -21.322 & -21.783 & -22.270 & -22.765 & -23.264 & -23.763 \\
\enddata
\end{deluxetable}
\clearpage

\begin{deluxetable}{llllllllllllll}
\scriptsize
\tablewidth{0pc}
\tablenum{7}
\tablecaption{O IV cooling}
\tablehead{}
\startdata
 2.0 & -20.687 & -20.693 & -20.712 & -20.768 & -20.909 & -21.182 & -21.575 & -22.036 & -22.522 & -23.018 & -23.516 & -24.016 & -24.516 \\
 2.1 & -20.279 & -20.284 & -20.301 & -20.348 & -20.470 & -20.720 & -21.097 & -21.550 & -22.034 & -22.529 & -23.027 & -23.527 & -24.026 \\
 2.2 & -19.974 & -19.979 & -19.992 & -20.033 & -20.140 & -20.369 & -20.730 & -21.175 & -21.656 & -22.150 & -22.648 & -23.147 & -23.647 \\
 2.3 & -19.750 & -19.754 & -19.766 & -19.801 & -19.897 & -20.108 & -20.454 & -20.891 & -21.369 & -21.862 & -22.360 & -22.859 & -23.359 \\
 2.4 & -19.591 & -19.594 & -19.605 & -19.636 & -19.723 & -19.919 & -20.252 & -20.682 & -21.157 & -21.649 & -22.146 & -22.645 & -23.145 \\
 2.5 & -19.483 & -19.486 & -19.495 & -19.524 & -19.603 & -19.787 & -20.107 & -20.531 & -21.003 & -21.494 & -21.992 & -22.491 & -22.990 \\
 2.6 & -19.416 & -19.418 & -19.427 & -19.453 & -19.525 & -19.698 & -20.007 & -20.424 & -20.894 & -21.384 & -21.881 & -22.380 & -22.879 \\
 2.7 & -19.381 & -19.383 & -19.391 & -19.414 & -19.481 & -19.642 & -19.940 & -20.349 & -20.816 & -21.305 & -21.801 & -22.300 & -22.800 \\
\enddata
\end{deluxetable}
\clearpage

\begin{deluxetable}{llllllllllllll}
\scriptsize
\tablewidth{0pc}
\tablenum{8}
\tablecaption{Ne III cooling}
\tablehead{}
\startdata
 2.0 & -23.949 & -23.949 & -23.949 & -23.949 & -23.950 & -23.952 & -23.960 & -23.983 & -24.048 & -24.206 & -24.500 & -24.907 & -25.373 \\
 2.1 & -23.169 & -23.169 & -23.169 & -23.169 & -23.170 & -23.172 & -23.179 & -23.200 & -23.260 & -23.407 & -23.688 & -24.086 & -24.548 \\
 2.2 & -22.557 & -22.557 & -22.557 & -22.557 & -22.558 & -22.560 & -22.567 & -22.586 & -22.642 & -22.779 & -23.046 & -23.436 & -23.894 \\
 2.3 & -22.078 & -22.078 & -22.078 & -22.078 & -22.079 & -22.081 & -22.087 & -22.106 & -22.157 & -22.285 & -22.541 & -22.922 & -23.377 \\
 2.4 & -21.704 & -21.704 & -21.704 & -21.704 & -21.705 & -21.707 & -21.713 & -21.731 & -21.779 & -21.900 & -22.145 & -22.518 & -22.968 \\
 2.5 & -21.414 & -21.414 & -21.414 & -21.414 & -21.415 & -21.416 & -21.422 & -21.440 & -21.487 & -21.602 & -21.837 & -22.202 & -22.649 \\
 2.6 & -21.191 & -21.191 & -21.191 & -21.191 & -21.192 & -21.194 & -21.200 & -21.217 & -21.262 & -21.373 & -21.600 & -21.958 & -22.401 \\
 2.7 & -21.022 & -21.022 & -21.022 & -21.022 & -21.023 & -21.025 & -21.031 & -21.047 & -21.092 & -21.199 & -21.420 & -21.772 & -22.212 \\
\enddata
\end{deluxetable}
\clearpage

\begin{deluxetable}{llllllllllllll}
\scriptsize
\tablewidth{0pc}
\tablenum{9}
\tablecaption{Ne IV cooling}
\tablehead{}
\startdata
 3.5 & -26.507 & -26.507 & -26.507 & -26.507 & -26.508 & -26.510 & -26.518 & -26.538 & -26.580 & -26.660 & -26.817 & -27.100 & -27.499 \\
 3.6 & -24.880 & -24.880 & -24.881 & -24.881 & -24.882 & -24.884 & -24.891 & -24.909 & -24.948 & -25.023 & -25.169 & -25.438 & -25.826 \\
 3.7 & -23.599 & -23.599 & -23.599 & -23.599 & -23.600 & -23.602 & -23.608 & -23.624 & -23.661 & -23.730 & -23.866 & -24.120 & -24.491 \\
 3.8 & -22.590 & -22.590 & -22.590 & -22.590 & -22.591 & -22.593 & -22.598 & -22.613 & -22.647 & -22.711 & -22.836 & -23.070 & -23.414 \\
 3.9 & -21.797 & -21.797 & -21.797 & -21.797 & -21.798 & -21.799 & -21.804 & -21.818 & -21.848 & -21.907 & -22.020 & -22.230 & -22.532 \\
 4.0 & -21.174 & -21.174 & -21.174 & -21.174 & -21.175 & -21.176 & -21.180 & -21.192 & -21.220 & -21.273 & -21.372 & -21.554 & -21.805 \\
 4.1 & -20.685 & -20.685 & -20.685 & -20.686 & -20.686 & -20.687 & -20.691 & -20.701 & -20.726 & -20.773 & -20.859 & -21.011 & -21.210 \\
 4.2 & -20.302 & -20.302 & -20.302 & -20.303 & -20.303 & -20.304 & -20.307 & -20.316 & -20.338 & -20.380 & -20.453 & -20.579 & -20.735 \\
\enddata
\end{deluxetable}
\clearpage

\begin{deluxetable}{llllllllllllll}
\scriptsize
\tablewidth{0pc}
\tablenum{10}
\tablecaption{Ne V cooling}
\tablehead{}
\startdata
 2.0 & -20.926 & -20.928 & -20.932 & -20.945 & -20.984 & -21.089 & -21.313 & -21.670 & -22.113 & -22.593 & -23.086 & -23.584 & -24.084 \\
 2.1 & -20.478 & -20.479 & -20.483 & -20.494 & -20.527 & -20.618 & -20.822 & -21.161 & -21.594 & -22.071 & -22.564 & -23.061 & -23.560 \\
 2.2 & -20.136 & -20.137 & -20.140 & -20.149 & -20.178 & -20.258 & -20.443 & -20.766 & -21.191 & -21.664 & -22.155 & -22.652 & -23.151 \\
 2.3 & -19.871 & -19.872 & -19.874 & -19.883 & -19.908 & -19.979 & -20.149 & -20.457 & -20.874 & -21.344 & -21.834 & -22.330 & -22.829 \\
 2.4 & -19.659 & -19.659 & -19.662 & -19.669 & -19.691 & -19.754 & -19.911 & -20.206 & -20.615 & -21.082 & -21.571 & -22.067 & -22.566 \\
 2.5 & -19.480 & -19.481 & -19.483 & -19.489 & -19.508 & -19.564 & -19.709 & -19.990 & -20.392 & -20.855 & -21.344 & -21.840 & -22.338 \\
 2.6 & -19.327 & -19.327 & -19.329 & -19.335 & -19.352 & -19.402 & -19.535 & -19.804 & -20.197 & -20.658 & -21.145 & -21.641 & -22.139 \\
 2.7 & -19.199 & -19.199 & -19.201 & -19.206 & -19.221 & -19.268 & -19.392 & -19.649 & -20.034 & -20.491 & -20.977 & -21.472 & -21.971 \\
\enddata
\end{deluxetable}
\clearpage

\begin{deluxetable}{lllllll}
\scriptsize
\tablewidth{0pc}
\tablenum{11}
\tablecaption{Cooling from other ions}
\tablehead{}
\tablehead{\colhead{} & \colhead{H~I} & \colhead{H~II} 
&\colhead{C~IV} & \colhead{O~V}}
\startdata
 3.5 & -33.939 & -22.684 & -29.877 & -48.346 \nl
 3.6 & -30.636 & -22.752 & -27.297 & -30.612 \nl
 3.7 & -28.019 & -22.820 & -25.257 & -28.029 \nl
 3.8 & -25.946 & -22.887 & -23.646 & -25.991 \nl
 3.9 & -24.305 & -22.954 & -22.375 & -24.386 \nl
 4.0 & -23.009 & -23.020 & -21.374 & -23.125 \nl
 4.1 & -21.985 & -23.085 & -20.588 & -22.136 \nl
 4.2 & -21.177 & -23.149 & -19.973 & -21.362 \nl
 4.3 & -20.542 & -23.212 & -19.494 & -20.753 \nl
 4.4 & -20.044 & -23.274 & -19.122 & -20.262 \nl
 4.5 & -19.654 & -23.334 & -18.836 & -19.849 \nl
 4.6 & -19.351 & -23.393 & -18.617 & -19.487 \nl
 4.7 & -19.116 & -23.450 & -18.453 & -19.169 \nl
 4.8 & -18.936 & -23.505 & -18.331 & -18.897 \nl
 4.9 & -18.799 & -23.559 & -18.244 & -18.673 \nl
 5.0 & -18.696 & -23.610 & -18.183 & -18.495 \nl
 5.1 & -18.621 & -23.660 & -18.144 & -18.357 \nl
 5.2 & -18.567 & -23.708 & -18.123 & -18.253 \nl
 5.3 & -18.531 & -23.754 & -18.114 & -18.178 \nl
 5.4 & -18.508 & -23.799 & -18.117 & -18.127 \nl
 5.5 & -18.496 & -23.843 & -18.128 & -18.095 \nl
 5.6 & -18.493 & -23.887 & -18.145 & -18.078 \nl
 5.7 & -18.496 & -23.929 & -18.169 & -18.073 \nl
 5.8 & -18.505 & -23.970 & -18.196 & -18.079 \nl
 5.9 & -18.518 & -24.009 & -18.227 & -18.092 \nl
 6.0 & -18.535 & -24.044 & -18.260 & -18.112 \nl
\enddata
\end{deluxetable}

\clearpage

\end{document}